\documentclass[10pt, conference]{IEEEtran}

\usepackage{graphicx}
\usepackage{algorithm}
\usepackage[noend]{algpseudocode}
\usepackage{textcomp}

\usepackage{subcaption}
\usepackage{url}
\usepackage{graphicx}

\usepackage{epsfig}
\usepackage{url}
\usepackage{stfloats}
\usepackage{enumerate}
\usepackage{longtable}

\usepackage{latexsym}
\usepackage{verbatim}
\usepackage{amssymb}

\usepackage{xspace}

\usepackage{xcolor}

\usepackage{verbatim}
\usepackage{caption}
\usepackage{subcaption}
\usepackage{amsmath}
\usepackage{amssymb}
\usepackage{wrapfig}
\usepackage{tabularx}
\usepackage{lscape}
\usepackage{float}
\usepackage{inputenc}
\usepackage[utf8]{luainputenc}
\usepackage[T1]{fontenc}
\usepackage{pgfgantt}
\usepackage{footnote}

\usepackage{algpseudocode}
\usepackage{algorithm}
\usepackage[para,online,flushleft]{threeparttable}
\usepackage{multicol}
\usepackage{multirow}
\usepackage{tablefootnote}
\usepackage{mathtools}
\usepackage{graphicx}
\usepackage{caption}
\usepackage{subcaption}
\usepackage{amsmath}
\usepackage{tikz}
\usepackage{float}
\usepackage{bm}
\usepackage{url}

\newcommand{\ignore}[1]{}
\usetikzlibrary{matrix,shapes,arrows,positioning,chains, calc}

\mathchardef\mhyphen="2D

\usepackage{pbox}

\newcommand{\algrule}[1][.2pt]{\par\vskip.5\baselineskip\hrule height #1\par\vskip.5\baselineskip}

\newcommand{\Rn}{\ensuremath \stackrel{\$}{\leftarrow}\mathbb{Z}_{n}^{*}{\xspace}}

\mathchardef\mhyphen="2D

\newcommand{\as}{\ensuremath {\leftarrow}{\xspace}}

{\xspace}

\newcommand{\bpvschkg}{\ensuremath {\mathit{BPV}}\mhyphen{\mathit{FourQ}}\mhyphen{\mathit{Schnorr.Kg}}{\xspace}}
\newcommand{\bpvschsig}{\ensuremath{\mathit{BPV}}\mhyphen{\mathit{FourQ}}\mhyphen{\mathit{Schnorr.Sig}}{\xspace}}
\newcommand{\bpvschver}{\ensuremath{\mathit{BPV}}\mhyphen{\mathit{FourQ}}\mhyphen{\mathit{Schnorr.Ver}}{\xspace}}

\newcommand{\bpvieskg}{\ensuremath{\mathit{BPV}}\mhyphen{\mathit{FourQ}}\mhyphen{\mathit{ECIES.Kg}}{\xspace}}
\newcommand{\bpviessig}{\ensuremath{\mathit{BPV}}\mhyphen{\mathit{FourQ}}\mhyphen{\mathit{ECIES.Enc}}{\xspace}}
\newcommand{\bpviesver}{\ensuremath{\mathit{BPV}}\mhyphen{\mathit{FourQ}}\mhyphen{\mathit{ECIES.Dec}}{\xspace}}

\newcommand{\params}{\ensuremath {\mathit{params}}{\xspace}}

\newcommand{\BPV}{\ensuremath {\mathit{BPV}}{\xspace}}
\newcommand{\BPVOff}{\ensuremath {\mathit{BPV.Offline}}{\xspace}}
\newcommand{\BPVOn}{\ensuremath {\mathit{BPV.Online}}{\xspace}}

\newcommand{\eat}[1]{}                

\newcounter{linecounter}

\usepackage[hang,flushmargin]{footmisc}

\usepackage{pifont}
\usepackage{wrapfig}

\IEEEoverridecommandlockouts

\newcommand\myfigure[5]{%
	\ifdim#2>.8\linewidth
	{%
		\centering
		\includegraphics[width=#3]{#4}%
		\captionof{figure}{#5}%
	}%
	\else
	\begin{wrapfigure}{#1}{#2}
		\includegraphics[width=#3]{#4}
		\caption{#5}
	\end{wrapfigure}
	\fi
}

\newcommand\blfootnote[1]{%
	\begingroup
	\renewcommand\thefootnote{}\footnote{#1}%
	\addtocounter{footnote}{-1}%
	\endgroup
}

\begin{document}

\title{Dronecrypt - An Efficient Cryptographic Framework for Small Aerial Drones}

\author{\IEEEauthorblockN{Muslum Ozgur Ozmen}
	\IEEEauthorblockA{Oregon State University\\
		Corvallis, Oregon, USA \\
		ozmenmu@oregonstate.edu}
	\and
	\IEEEauthorblockN{Attila A. Yavuz}\thanks{Work done in part while Attila A. Yavuz was at Oregon State University, Corvallis, OR.}
	\IEEEauthorblockA{University of South Florida\\
		Tampa, FL, USA \\
		attilaayavuz@usf.edu}}

\maketitle

\begin{abstract}
Aerial drones are becoming an integral part of application domains including but not limited to, military operations, package delivery, construction, monitoring and search/rescue operations. It is critical to ensure the cyber security of networked aerial drone systems in these applications. Standard cryptographic services can be deployed to provide basic security services; however, they have been shown to be inefficient in terms of energy and time consumption, especially for small aerial drones with resource-limited processors. Therefore, there is a significant need for an efficient cryptographic framework that can meet the requirements of small aerial drones.

{\em We propose an improved cryptographic framework for small aerial drones, which offers significant energy efficiency and speed advantages over standard cryptographic techniques}. (i) We create (to the best of our knowledge) the first optimized public key infrastructure (PKI) based framework for small aerial drones, which provides energy efficient techniques by harnessing special precomputation methods and optimized elliptic curves. (ii) We also integrate recent light-weight symmetric primitives into our PKI techniques to provide a full-fledged cryptographic framework. (iii) We implemented standard counterparts and our proposed techniques on an actual small aerial drone (Crazyflie 2.0), and provided an in-depth energy analysis. Our experiments showed that {\em our improved cryptographic framework achieves up to 35$\times$ lower energy consumption than its standard counterpart}. 

\blfootnote{$\copyright$ 2019 IEEE. Personal use of this material is permitted. Permission from IEEE must be obtained for all other uses, in any current or future media, including reprinting/republishing this material for advertising or promotional purposes, creating new collective works, for resale or redistribution to servers or lists, or reuse of any copyrighted component of this work in other works.}
\end{abstract}

\IEEEpeerreviewmaketitle

\section{Introduction}\label{sec:Introduction}

\begin{figure*}[!t]
	\centering
	\includegraphics[width=\linewidth]{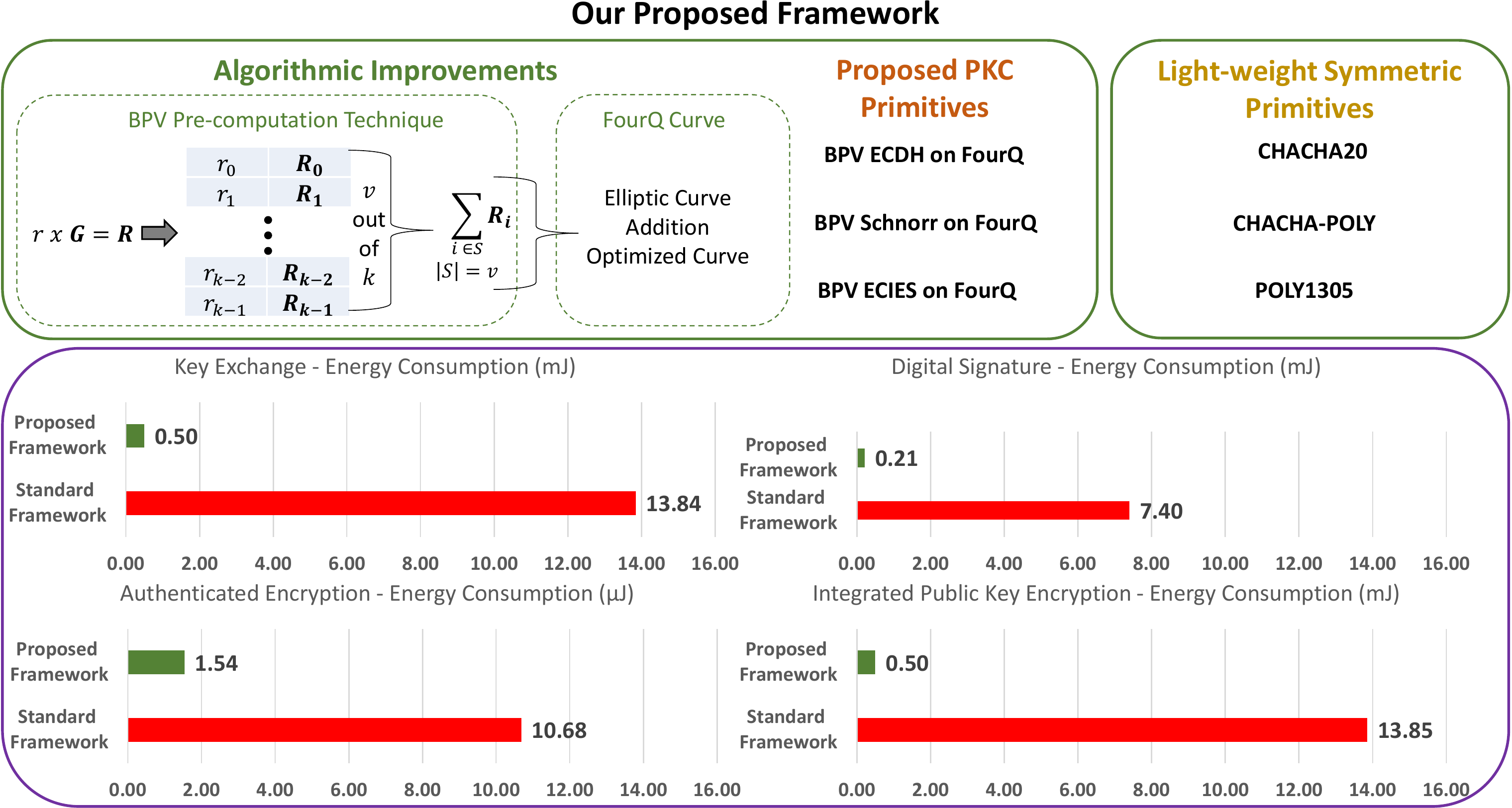}
	\caption{The performance comparison of the proposed and standard cryptographic frameworks.}
	\label{fig:intro}
\end{figure*}

Aerial drones are emerging mobile cyber-physical systems with potential applications such as military operations, package delivery, reconnaissance, environmental monitoring and disaster recovery/response~\cite{DroneUseCases1, DroneDisaster,DroneUseCase::Positioning, DRONE:Military}. Due to the significant financial and strategic value involved, aerial drones are expected to be exclusively targeted by attackers. These attacks are especially critical for military networks where some sensitive information can be extracted. For instance, in 4th December 2011, a US drone (RQ-170 Sentinel) was captured by Iranian forces. A strong theory that explains how it was captured supposes that the navigation system of the drone was attacked~\cite{DRONE:Military, Drone:Security:Survey}. Another incident is the keylogging attack that targeted a US drone fleet in September 2011~\cite{DRONE:Military}. This attack can cause the exposure of the control station and potentially result in the full control on the drone. 

Therefore, it is critical to ensure the cyber security of aerial drone systems. In particular, the basic security services such as confidentiality, authentication, and integrity must be provided to refrain from mentioned attacks. These services are mainly guaranteed via fundamental cryptographic techniques such as symmetric ciphers and PKI. Symmetric key cryptography is highly preferred due to its efficiency and well-analyzed security. However, public key primitives can also be highly useful for certain drone applications. For instance, public key cryptography offers scalability, public verifiability, and non-repudiation that are critical for large and distributed drone networks (e.g., military operations, search and rescue).

Energy and bandwidth constraints pose critical limitations towards the deployment of standard cryptographic techniques on small aerial drones~\cite{bitcraze2016crazyflie}. In the following, we outline recent cryptographic techniques that are considered for aerial drones and highlight some obstacles against practical deployments.

\subsection{Related Work} \label{subsec:RelatedWorkGap}

{\em Security Vulnerabilities and Solutions}: Son et al.~\cite{DroneSecurity::Yongdae} pointed out that most of commodity (civilian) drone telemetry systems do not provide a cryptographic protection to secure the communication. Thus, they proposed a fingerprinting method that would potentially provide some authentication for drones in motion. Some security vulnerabilities of current drone configurations were shown in~\cite{DroneSecurity::Yongdae2,DroneSecurity::Yongdae3}. That is, it might be possible to extract frequency hopping sequence of FHSS-type drone controllers using a software defined radio~\cite{DroneSecurity::Yongdae2}. Moreover, it is possible make a drone crash by exploiting the resonance frequency of MEMS gyroscopes~\cite{DroneSecurity::Yongdae3}. Birnbach et al.~\cite{DroneSecurity::Martinovic} identified the privacy invasion attacks performed by drones and offered detection mechanisms.

{\em Cryptographic Techniques}: Initial approaches to secure the aerial drones include implementation of well-known protocols (e.g., RSA and AES) on FPGAs~\cite{Drones:RSA/AES:FPGA, Drones:Crypto:Weight}. However, they showed that standard techniques take a lot of computation time and consume high amount of energy on small aerial drones that are further confirmed by our experiments (see Section \ref{sec:Performance}). 

Seo et al.~\cite{Drones:BertinoSeo:2016:Framework} proposed a security framework that implemented some symmetric ciphers with white-box cryptography to mitigate the impacts of drone capture attacks~\cite{Drones:BertinoSeo:2016:Framework}. However, the insecurity of this line of research has been later shown in~\cite{WhiteBoxAttack}. Won et al. proposed Certificateless Signcryption based protocols for mid-size aerial drones (e.g., AR.Drone 2.0) in~\cite{Drone:Elisa:Won:2017}. This protocol reduces the communication overhead by eliminating certificates, but requires several exponentiations and therefore introduces heavy overhead, which might not be practical for resource-limited small aerial drones. Moreover, aerial drones are expected to be an integral part of Internet of Things (IoT), which vastly relies on PKI technology. Certificateless protocols cannot be seamlessly integrated into existing PKI-based IoT systems without significant alterations.

Precomputations and optimizations of public key cryptography algorithms were considered on power constrained nodes. These include sole implementations of certain primitives (e.g., digital signature~\cite{BPV:Ateniese:Journal:ACMTransEmbeddedSys:2017}) and improvements on a few asymmetric primitives (e.g.,~\cite{Ozmen_IOT_SP}). However, these techniques do not harbor the optimized elliptic curves (e.g., FourQ~\cite{FourQBase}) with the precomputation techniques to offer a comprehensive framework.

One may observe that the following issues should be addressed towards a practical deployment of cryptographic techniques on small aerial drones (i) The existing standard primitives and cryptographic protocols are inefficient in terms of energy/time consumption for small aerial drones that operates with resource-limited microprocessors (see Figure~\ref{fig:intro}). (ii) The existing cryptographic protocols for aerial drones only focus on a few specific primitives but do not offer a comprehensive framework that harbors a variety of light-weight symmetric and asymmetric primitives. (iii) To the best of our knowledge, there is no open-source framework, even for some standard cryptographic primitives, which offers a {\em detailed energy assessment}.

\subsection{Our Contribution} \label{subsec:Contributions}

Towards the practical adoption of cryptographic techniques to resource-limited small aerial drones, we propose an improved cryptographic framework by harnessing various cryptographic primitives and optimizations, which provides significantly lower delay and energy consumption for small aerial drones, compared to the deployment of standard cryptographic techniques. We further outline our contributions as follows:

$\bullet$ {\em An Efficient Framework with Algorithmic Improvements}: Our goal is to reduce the overhead of cryptographic primitives to minimize their energy consumption for small aerial drones. Therefore, we exploit synergies among special precomputation techniques and elliptic curves (EC), which not only offer the most compact key/signature sizes among available alternatives, but also significantly reduce the computational cost of EC scalar multiplication.

We develop (to the best of our knowledge) the first realization of Boyko-Peinado-Venkatesan (BPV) precomputation technique~\cite{BPV:basepaper:1998} on FourQ curve~\cite{FourQBase}. BPV technique reduces the cost of an EC scalar multiplication to only a few EC additions with only a small constant-size storage overhead. Remark that FourQ curve is one of the most addition friendly curves (even with a better efficiency than that of Curve25519~\cite{Curve25519Base}), and therefore our integration further enhances the efficiency of BPV. We then instantiate {\em BPV-FourQ-Schnorr} and {\em BPV-FourQ-ECIES} as our improved digital signature and integrated encryption schemes, which significantly outperform their standard counterparts. Moreover, we also integrate some of the recent light-weight symmetric primitives into our improved PKI suite to create a full-fledged cryptographic framework.

$\bullet$ {\em In-depth Energy Analysis}: The energy consumption of standard Elliptic Curve Cryptography (ECC) based primitives have not been investigated thoroughly for recently emerging small aerial drones. In this paper, we implemented both standard ECC techniques and our proposed cryptographic framework and presented a detailed energy consumption analysis. As demonstrated in Figure~\ref{fig:intro}, for different cryptographic primitives, our experiments showed that proposed improvements enable up to 35$\times$ less energy consumption compared to the standard techniques for small aerial drones (see Section~\ref{sec:Performance}). Similar performance gains were observed for the light-weight ciphers over the standard symmetric primitives.

$\bullet$ {\em Open-Source Framework}: While isolated implementation results were reported for particular cryptographic primitives such as RSA and AES~\cite{Drones:RSA/AES:FPGA, Drones:Crypto:Weight}, to the best of our knowledge, no comprehensive open-source cryptographic framework is available for small aerial drones. Towards meeting this need, we implemented our standard and improved cryptographic frameworks. We open-source both to enable a broad test and potential adoption at 

\begin{center}
	\url{https://github.com/ozgurozmen/Dronecrypt}
\end{center}

\noindent \textbf{Limitations: }The main limitation of our proposed framework is the increased private key size, that is inherited from the BPV precomputation technique. More specifically, our improved PKI primitives require the sender to store a private key of 64 KB. However, this increased private key size translates into significant improvements on computational time and energy consumption. Moreover, only a small portion of this key (1 KB) must be loaded into RAM for computations. Considering the storage capabilities of small aerial drones, we believe this is a desirable trade-off for security-critical applications. On the other hand, if the drone is highly memory-limited and cannot tolerate such storage, standard cryptographic primitives should be preferred.

\section{Preliminaries}\label{sec:Prelim}

\begin{table}[h]
	\centering
	\caption{Notation followed to describe schemes.} \label{tab:Notation}
	\begin{threeparttable}
		\begin{tabular}{| c | c |}
			\hline
			$F_q$ & Finite Field \\ \hline
			$\mathbf{G}$ & Generator Group Point \\ \hline
			$n$ & Order of Group \\ \hline
			$\langle y,\mathbf{Y}\rangle$ & Private/Public key pair \\ \hline
			$\Gamma$ & BPV Precomputation Table \\ \hline
			$m$ & Message  \\ \hline
			$\mathcal{E}_{k}$ & IND-CPA Encryption via key $k$ \\ \hline
			$\mathcal{D}_{k}$ & IND-CPA Decryption via key $k$ \\ \hline
			$\times$ & Elliptic Curve Scalar Multiplication (Emul) \\ \hline
 			$KDF$ & Key Derivation Function \\ \hline
		\end{tabular}
		Elliptic Curve (EC) points are shown in bold
	\end{threeparttable}
\end{table}

\noindent \textbf{BPV Pre-computation Technique: }We use BPV~generator~\cite{BPV:basepaper:1998}, which reduces the computational cost of an Emul to a few EC additions with the expense of a table storage. The \BPV~generator is a tuple of two algorithms $(\mathit{Offline},\mathit{Online})$ defined in Algorithm~\ref{alg:BPV}. \\ \vspace{-2mm}

\begin{algorithm}[h!]
	\caption{Boyko-Peinado-Venkatesan (BPV) Generator}\label{alg:BPV}
	\hspace{5pt}
	\begin{algorithmic}[1]
		\Statex $\underline{(\Gamma,v,k,F_q,\mathbf{G},n) \as \BPVOff(1^{\kappa})}$: 
		\vspace{3pt}
		\State Set the EC system-wide parameters $\params\as(F_q,\mathbf{G},n)$.
		\State Generate \BPV~parameters $(v,k)$, where $k$ and $v$ are the number of pairs to be pre-computed and the number of elements to be randomly selected out $k$ pairs, respectively, for $2<v<k$.
		\State $r'_i\Rn,$~~ $\mathbf{R'_i} \as {r'_i} \times \mathbf{G}$, $i=0,\ldots, k-1$.
		\State Set pre-computation table $\Gamma=\{r'_i,\mathbf{R'_i}\}_{i=0}^{k-1}$.
	\end{algorithmic}
	\algrule
	\begin{algorithmic}[1]
		\Statex $\underline{(r,\mathbf{R}) \as \BPVOn(\Gamma,v,k,F_q,\mathbf{G},n)}$:
		\vspace{3pt}
		\State Generate a random set $S \subset [0,k-1]$, where $|S| = v$.
		\State $r \as \sum_{i \in S}^{} r'_i \bmod n$, $\mathbf{R} \as \sum_{i \in S}^{} \mathbf{R'_i}$.
	\end{algorithmic}
	
\end{algorithm}

\noindent \textbf{FourQ Curve: }FourQ is a high-security, high-performance elliptic curve~\cite{FourQBase}. FourQ synergizes some well-known EC optimizations to offer high-speed EC scalar multiplication and EC addition while preserving 128-bit security. \\ \vspace{-3mm}

\noindent \textbf{Standard Techniques: } Our standard cryptographic framework consists of broadly standardized techniques.

We select secp256k1 curve to implement our standard public key cryptography services, such as ECDH~\cite{DH}, ECDSA~\cite{ECDSA}, and ECIES~\cite{Pointcheval00psec3}. Secp256k1 is a NIST recommended curve~\cite{NIST_ECParameters}, which is frequently used in practice. We implemented a key exchange, a digital signature and an integrated scheme to secure a variety of applications. We also implemented some of the most well-known symmetric key cryptography techniques such as AES as the block cipher~\cite{AES}, AES-GCM as the authenticated encryption~\cite{AESGCM} and HMAC with SHA-256 for MAC. Note that, an open-source implementation and in-depth energy analysis of such standard cryptographic framework for small aerial drones are not currently available (to the best of our knowledge). 

\section{Proposed Framework}\label{sec:proposedschemes}%

\subsection{Low Cost PKC Primitives}
One may consider adopting traditional pre-computation methods to reduce the computation overhead of cryptographic techniques. However, traditional online/offline pre-computation methods require linear storage. Thus, they may not be feasible for small aerial drones equipped with light-weight microcontrollers. In addition, once the pre-computed tokens are depleted, they must be regenerated. It is shown that the regeneration cost of these pre-computation techniques may incur even more cost than following the original protocol~\cite{MILCOM:Yavuz}. 

We propose an integrated approach that only requires small-constant storage overhead and highly optimized EC addition operations as opposed to a full EC scalar multiplication. To achieve this objective, we present an instantiation of BPV on FourQ curve, which is ideal for BPV due to its very fast EC additions. With BPV, we reduce EC scalar multiplication operations of our target schemes/protocols to only a few EC addition operations. The integration of FourQ into BPV amplifies the performance gain as it is an addition efficient curve. As it will be showcased in Section \ref{sec:Performance}, this strategy enables almost 35$\times$ more efficient operations compared to the standard schemes (with standard curves such as secp256k1). The storage cost of such an energy efficiency gain is only a constant-size storage of 64KB (for parameters $v = 16, k = 1024$, see Section~\ref{sec:Prelim}) keying material. With the current capabilities of low-end processors, even on the small aerial drones (such as Crazyflie 2.0~\cite{bitcraze2016crazyflie}), it is feasible to store such a private key. We then use BPV-FourQ to instantiate three main cryptographic schemes, which creates the following efficient digital signature, key exchange, and integrated encryption suite:

(i) We integrate our optimized BPV-FourQ into Schnorr signature scheme to gain computational efficiency. This transformation is presented in Algorithm \ref{alg:BPV-Schnorr}.

\begin{algorithm}[h!]
	\caption{BPV-FourQ-Schnorr Signature}\label{alg:BPV-Schnorr}
	\hspace{5pt}
	\begin{algorithmic}[1]
		\Statex $\underline{(\Gamma,y,\mathbf{Y}) \leftarrow \bpvschkg(1^{\kappa})}$: 
		\vspace{3pt}
		\State Generate parameters and BPV table as $(\Gamma, v,k,F_q,\mathbf{G},n) \as \BPVOff(1^{\kappa})$.
		\State Generate private/public key pair $(y\stackrel{\$}{\leftarrow}\mathbb{Z}_{n}^{*},\mathbf{Y}\leftarrow
		y \times \mathbf{G} )$
	\end{algorithmic}
	\algrule
	
	\begin{algorithmic}[1]
		\Statex $\underline{(s,e)\leftarrow \bpvschsig(m,y,\Gamma)}$:
		\vspace{3pt}
		\State ${(r,\mathbf{R}) \as \BPVOn(\Gamma,v,k,F_q,\mathbf{G},n)}$
		\State $e\as H(m||\mathbf{R}),~~s\as (r-e\cdot y) \bmod n$ where $H$ is a full domain cryptographic hash function $H:\{0,1\}^{*} \rightarrow \mathbb{Z}_{n}^{*}$.
	\end{algorithmic}
	\algrule
	
	\begin{algorithmic}[1]
		\Statex $\underline{b\leftarrow \bpvschver(m,\langle s,e \rangle,\mathbf{Y})}$: 
		\vspace{3pt}
		\State $\mathbf{R'}\as  e \times \mathbf{Y} + s \times \mathbf{G}$
		\State If $e=H(m||\mathbf{R'})$ then  set $b=1$ as {\em valid}, else $b=0$ .
	\end{algorithmic}
\end{algorithm}

(ii) Similarly, we instantiate ECDH with BPV-FourQ. In the ECDH variant, each node first derives its private/public key pair with an EC scalar multiplication, and then one more EC scalar multiplication is performed in order to obtain the shared key~\cite{DH}. We adopt BPV protocol so that each node derives its private/public key pair with only EC additions. This decreases the computation time and energy consumption of the protocol by almost 1.5$\times$, in the expense of storing a 64KB table. Moreover, it is almost 28$\times$ more efficient than its standard counterpart (ECDH on secp256k1, see Section \ref{sec:Performance}).

(iii) In ECIES protocol, node 1 first derives the shared secret by generating an ephemeral private/public key pair and then computing the shared secret using node 2's public key~\cite{Pointcheval00psec3}. Then, she generates encryption and MAC keys from this shared secret with a pre-determined key derivation function. We adopt BPV to ECIES by transforming the first EC scalar multiplication (where the ephemeral private/public key pair is generated) to EC additions, as depicted in Algorithm~\ref{alg:BPV-ECIES}.

\begin{algorithm}[h!]
	\caption{BPV-FourQ-ECIES Encryption }\label{alg:BPV-ECIES}
	\hspace{5pt}
	\begin{algorithmic}[1]
		\Statex $\underline{(\Gamma,y,\mathbf{Y}) \leftarrow \bpvieskg(1^{\kappa})}$: 
		\vspace{3pt}
		\State Generate parameters and BPV table as $(\Gamma, v,k,F_q,\mathbf{G},n) \as \BPVOff(1^{\kappa})$.
		\State Generate private/public key pair $(y\stackrel{\$}{\leftarrow}\mathbb{Z}_{n}^{*},\mathbf{Y}\leftarrow
		y \times \mathbf{G})$
	\end{algorithmic}
	\algrule
	
	\begin{algorithmic}[1]
		\Statex $\underline{(c,d,\mathbf{R})\leftarrow \bpviessig (m,Y,\Gamma)}$:
		\vspace{3pt}
		\State ${(r,\mathbf{R}) \as \BPVOn(\Gamma,v,k,F_q,\mathbf{G},n)}$
		\State Generate shared secret as $\mathbf{T} \leftarrow r \times \mathbf{Y}$
		\State $(k_{enc}, k_{MAC}) \leftarrow KDF(\mathbf{T})$
		\State $c \leftarrow \mathcal{E}_{k_{enc}}(m)$
		\State $d \leftarrow MAC_{k_{MAC}}(c)$
	\end{algorithmic}
	\algrule
	
	\begin{algorithmic}[1]
		\Statex $\underline{m\leftarrow \bpviesver (y,\langle c,d,R \rangle)}$: 
		\vspace{3pt}
		\State $\mathbf{T'}\as y \times \mathbf{R} \bmod p$
		\State $(k_{enc}, k_{MAC}) \leftarrow KDF(\mathbf{T'})$
		\State If $d \neq  MAC_{k_{MAC}}(c)$ return INVALID
		\State $m \as \mathcal{D}_{k_{enc}}(c)$
	\end{algorithmic}
\end{algorithm}

\subsection{Low Cost Symmetric Key Primitives}

We integrate light-weight ciphers and symmetric authentication mechanisms into our framework. Specifically, we use CHACHA20 as a very fast stream cipher~\cite{BernsteinCHACHA}, CHACHA-POLY as the authenticated encryption and POLY1305 as the MAC protocol~\cite{rfc7539}. These schemes provide faster and energy-efficient encryption/authentication while still offering high-security guarantees~\cite{CHACHAPOLYSecurity}. Our experiments confirmed that the adoption of these light-weight ciphers offer significant improvements in terms of computation time and energy consumption. For instance, these light-weight symmetric techniques enable up-to 7$\times$ improvement over standard symmetric ciphers (e.g., AES) for small aerial drones (see Section~\ref{sec:Performance}). 

\section{Security of the Proposed Framework}

The security of our improved PKC suite relies on the security of the optimization techniques (BPV precomputation technique and FourQ curve), since the underlying schemes (Schnorr, ECDH and ECIES) are all well-studied and standardized. BPV technique takes advantage of the random walks on Cayley graphs over Abelian groups and it relies on the hardness of hidden subset sum problem. Boyko et al, show that any attack to BPV technique would require solving non-linear lattice problems~\cite{BPV:basepaper:1998}. BPV technique in elliptic curve based schemes (as adopted in our proposed framework) were later investigated in~\cite{BPV:Ateniese:Journal:ACMTransEmbeddedSys:2017}, with an integration to ECDSA. This integration relies on the affine hidden subset sum problem. Since the BPV technique is used in our schemes without any modification, it inherits its security guarantees presented in~\cite{BPV:basepaper:1998, BPV:Ateniese:Journal:ACMTransEmbeddedSys:2017}. As discussed in Section~\ref{sec:Prelim}, FourQ is a special elliptic curve that combines multiple optimizations to enable fast operations. While offering high computational efficiency, FourQ still preserves 128-bit security level that is the same with standardized secp256k1 curve~\cite{FourQBase}.

The security of the light-weight ciphers and symmetric authentication mechanisms are studied in~\cite{CHACHAPOLYSecurity}. Their security is equivalent to their well-known standard counterparts such as AES. Moreover, we used CHACHA20, that is the 20 round CHACHA scheme to ensure its security~\cite{CHACHAPOLYSecurity}. Therefore, the security of our proposed framework is well-analyzed and comparable to the standard framework.

\section{Performance Evaluation}\label{sec:Performance}

\subsection{Experimental Setup and Evaluation Metrics}
We worked on Crazyflie 2.0 (Figure 2) due to its open-source software and hardware~\cite{bitcraze2016crazyflie}. Crazyflie 2.0 has two microcontrollers: (i) An STM32F405 microcontroller as the main controller that runs all the flight control codes. Both cryptographic frameworks were implemented on this microcontroller due to its capabilities. (ii) An ultra light-weight nRF51822 microcontroller is responsible for the communication (radio) and power management. 

\myfigure{R}{.45\linewidth}{\linewidth}{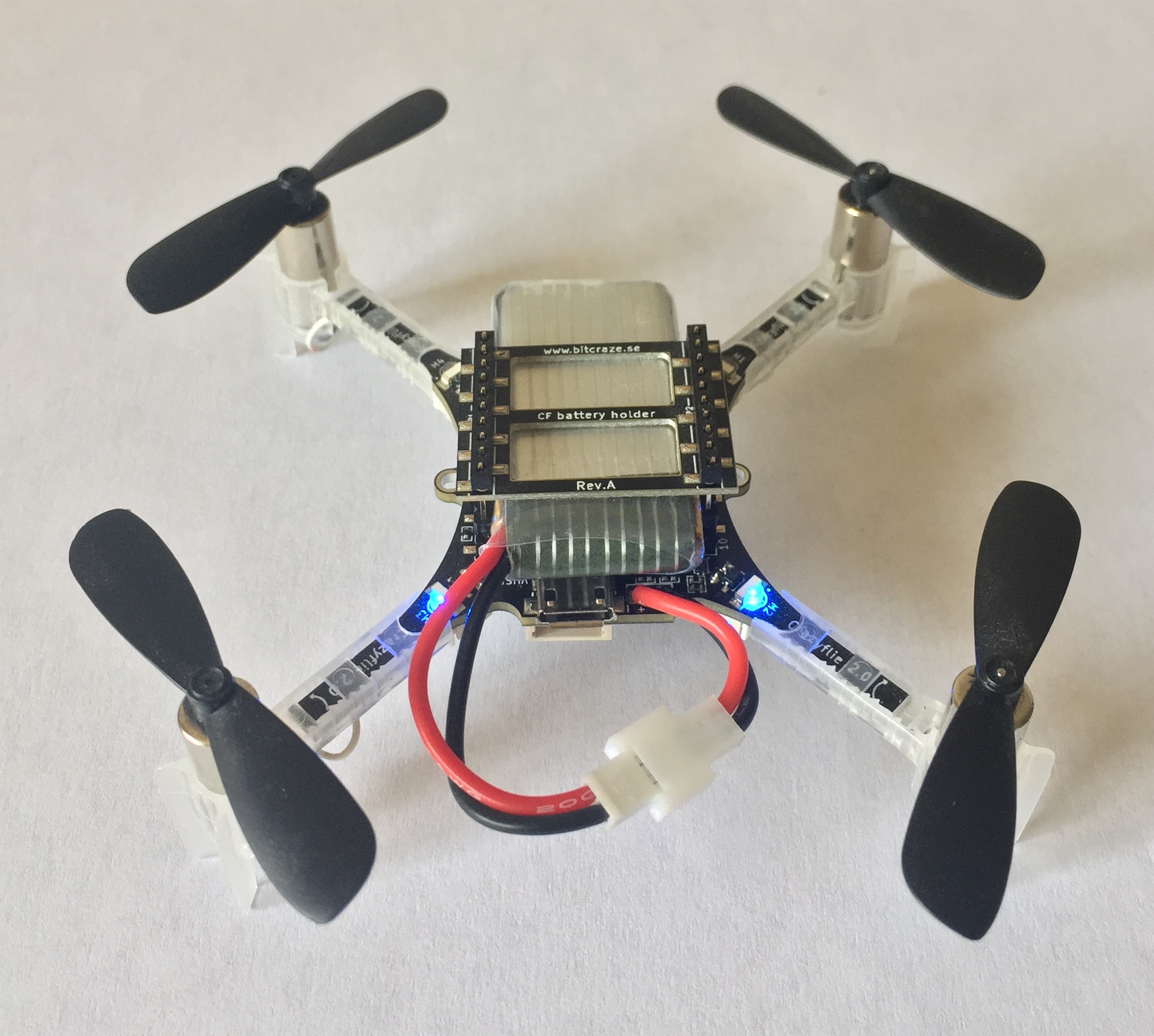}{Crazyflie 2.0}

STM32F405 is equipped with an ARM Cortex M-4 architecture and operates at 168 MHz. It is a 32-bit microcontroller with a 192 KB SRAM and 1 MB flash memory. Although STM32F405 is a resourceful processor, it has very low power consumption. It operates at $3.3V$ and takes $40mA$ of current while operating at 168 MHz.
Our evaluation metrics include computation time, storage, communication bandwidth, and energy consumption. We measured the energy consumption with the formula $E = V \cdot I \cdot t$, where $V = 3.3V$, $I = 40mA$~\cite{STM32F405Energy} and $t$ is the computation time based on the clock cycles. Moreover, we connected an ammeter between the device and the battery to double check the current taken by the processor and observed insignificant difference with $40 mA$, while running at 168 MHz.

We used the following libraries in our implementations. (i) All symmetric primitives are implemented using WolfCrypt~\cite{WolfCrypt}. (ii) We implemented all standard public key services using microECC~\cite{mackaymicro} as it is a light-weight and open-source library. (iii) All optimized public key primitives are implemented with open-source Microsoft FourQ library~\cite{FourQBase}. We selected BPV parameters as $v = 16$ and $k = 1024$, which sets the size of $\Gamma$ as 64 KB, that can be easily stored in STM32F405. We note that all libraries used for both standard and optimized frameworks are open-source, therefore they can be easily integrated to various aerial drones.

\begin{table*}[t!]
	\centering
	\caption{Comparison of Standard and Optimized Public Key Cryptographic Frameworks} \label{tab:PublicKey}
	\vspace{-2mm}
		\begin{threeparttable}
			\begin{tabular}{| c || c | c |  c | c | c | }
				\hline
				\textbf{Protocol} & \textbf{CPU Cycles} & \textbf{CPU Time (ms)} & \textbf{Memory\textsuperscript{$ \mathparagraph$} (Byte)} & \textbf{Bandwidth (Byte)} & \textbf{Energy Consumption (mJ)} \\ \hline
				
				\multicolumn{6}{|c|}{\textbf{\em Standard Cryptographic Framework}} \\ \hline \hline
				
				ECDH & 17611413 & 104.83 & 32 & 32 & 13.84 \\
				
				ECDSA-Sign & 9411839 & 56.02 & 32 & 64 & 7.40 \\
				
				ECDSA-Verify & 8169117 & 48.62 & 32 & 64 & 6.42 \\
				
				ECIES-Encrypt & 17625012 & 104.91 & 32 & 32 + |c| + |MAC| & 13.85 \\
				
				ECIES-Decrypt & 8817657 & 52.49 & 32 & 32 + |c| + |MAC| & 6.93 \\ \hline
		
				\multicolumn{6}{|c|}{\textbf{\em Optimized Cryptographic Framework with Algorithmic Improvements}} \\ \hline \hline
				
				BPV ECDH on FourQ & \textbf{636833} & \textbf{3.79} & 65536 & 32 & \textbf{0.50} \\
				
				BPV Schnorr-Sign on FourQ & \textbf{264011}  & \textbf{1.57} & 65536 & 64 & \textbf{0.21} \\
				
				BPV Schnorr-Verify on FourQ & \textbf{683882}  & \textbf{4.07} & 32 & 64 & \textbf{0.54} \\
				
				BPV ECIES-Encrypt on FourQ & \textbf{638791} & \textbf{3.80} & 65536 & 32 + |c| + |MAC|& \textbf{0.50} \\
				
				BPV ECIES-Decrypt on FourQ & \textbf{513487} & \textbf{3.06} & 32 & 32 + |c| + |MAC| & \textbf{0.40} \\ \hline
				
			\end{tabular}
			\begin{tablenotes}[flushleft] \scriptsize{
					$\mathparagraph $ Memory denotes the private key size for sign/encrypt schemes and the public key size for verify/decrypt schemes. {We want to note that even though the total memory required for BPV schemes is 64 KB (that includes $t = 1024$ components), only 1 KB ($v = 16$) of it is loaded into RAM for a single operation.}
				}
			\end{tablenotes}
		\end{threeparttable}
	\vspace{-1mm}
\end{table*}

\begin{table*}[t!]
	\centering
	\caption{Comparison of Standard and Optimized Symmetric Cryptographic Frameworks} \label{tab:Symmetric}
	\begin{threeparttable}
		\begin{tabular}{| c || c | c | c | c | }
			\hline
			\textbf{Protocol} & \textbf{(MB/s)} & \textbf{CPU Cycles\textsuperscript{$ \mathparagraph$}} & \textbf{CPU Time ($\mu$s)} &  \textbf{Energy Consumption ($\mu$J)} \\ \hline

			\multicolumn{5}{|c|}{\textbf{\em Standard Cryptographic Framework}} \\ \hline \hline
			
			AES & 0.926 & 5537 & 32.96 & 4.35\\
					
			AES-GCM & 0.377 & 13599 & 80.95 & 10.68 \\
			
			HMAC\tnote{$ \dagger $} & 3.338 & 1536 & 9.14 & 1.21 \\ \hline
			
			\multicolumn{5}{|c|}{\textbf{\em Optimized Cryptographic Framework}} \\ \hline \hline
			
			CHACHA20 & \textbf{3.554} & \textbf{1443} & \textbf{8.59} & \textbf{1.13} \\
			
			CHACHA-POLY & \textbf{2.619} &  \textbf{1958} & \textbf{11.65} & \textbf{1.54} \\
			
			POLY1305 & \textbf{13.709} & \textbf{374} & \textbf{2.23} &  \textbf{0.29} \\  \hline
			
		\end{tabular}
		\begin{tablenotes}[flushleft] \scriptsize{
				$\mathparagraph $ CPU cycles presented here are for a 32-byte message. \\
				$ \dagger $ SHA256 is used as the standard hash function for HMAC.
			}
		\end{tablenotes}
	\end{threeparttable}
\end{table*}

\subsection{Performance Evaluation and Comparison}

Experimental comparison of the standard framework and our proposed framework are presented in Tables~\ref{tab:PublicKey} and \ref{tab:Symmetric}, for public key and symmetric cryptography, respectively.

Although energy consumption is critical for small aerial drones, computation time also has some crucial effects on the adoption of cryptographic protocols in practice. For instance, there are time-critical applications that need frequent data transmission (e.g. camera mounted on an aerial drone for video surveillance - 24 frames per second are necessary). Cryptographic primitives used to secure such applications should meet this demand by offering high-speed operations. Therefore, we believe CPU Time improvements depicted in Tables~\ref{tab:PublicKey} and~\ref{tab:Symmetric} are also critical for small aerial drones.

Below, we present the comparison between the standard framework and our proposed framework, with an application to use-cases for small aerial drones.

$\bullet$~{\em {Digital Signature and Broadcast Authentication}}: Digital signatures are commonly used for broadcast authentication and key certification purposes. Drones may be used in scenarios where the drone needs to broadcast its sensor data, such as GPS data for location, photo frames for monitoring and temperature/pressure data for meteorological observation. Since digital signatures offer scalability, public verifiability, and non-repudiation properties that are lacked by symmetric key authentication mechanisms, they should be preferred for broadcast authentication. Moreover, certificates are necessary to protect key exchange schemes from man-in-the-middle attacks. Digital signatures are extensively used in practice for certification. Therefore, whenever a drone makes a key exchange with a server or another drone, certificate verification (digital signature verification) must be performed.

Our proposed framework offers significant improvements over the standard framework. As depicted in Figure~\ref{fig:intro}, energy consumption of digital signature is decreased to 0.21mJ from 7.40mJ with a 35.24$\times$ improvement. Moreover, the maximum signing throughput of the standard framework is 17 messages per second, that may not be sufficient for real-time use-cases.

$\bullet$~{\em {Integrated Public Key Encryption and Key Exchange}}: Although public key encryption is costly compared to symmetric key encryption, there are various use-cases that it can be useful. One of such cases could be when multiple drones need to report to a single server. We believe public key encryption can be useful for applications such as military drone fleets, tracking and search/rescue operations, due to the scalability concerns of symmetric key primitives. Moreover, ECIES provides forward security, which is useful for security-critical applications such as military operations.

A key exchange protocol is essential for aerial drone networks for the management and distribution of symmetric keys. That is, it might not be always possible to assume pre-installed symmetric keys for all drones as the controller/server may change for different purposes/use-cases. Therefore, an efficient key exchange protocol is useful to be deployed on drones.

Our optimized framework with algorithmic improvements achieves significantly lower energy consumption and faster encryption/key exchange than that of the standard framework. BPV-FourQ-ECIES is 27.70$\times$ more energy efficient than standard ECIES protocol. BPV-FourQ-ECDH also improved its standard counterpart by 27.68$\times$. Besides the energy efficiency, our optimized protocols offer fast computation that is important for time-critical applications.

$\bullet$~{\em {Light-weight Symmetric Primitives}}: Symmetric key cryptography can be used for various use-cases on drones, due to their low energy consumption. For instance, securing the command and control channel is the minimum requirement for a safe operation of any aerial drone system. This channel can be easily secured via symmetric key primitives. We suggest using an authenticated symmetric encryption to secure this communication channel. Moreover, when a single aerial drone reports its sensor data to one or a few base stations, symmetric key primitives can be preferred over public key schemes. 

Although the energy consumption of symmetric primitives is minimal compared to PKC, it is still useful to optimize their energy consumption as their use with high message throughputs might be much higher. For example, 100 messages per second are necessary to have a stable flight, which means 100 encryption/authentication operations per second~\cite{bitcraze2016crazyflie}. When authenticated encryption schemes are used to secure this channel as suggested, our proposed framework offers 6.95$\times$ lower energy consumption. Our proposed framework still achieves 3.84$\times$ and 4.11$\times$ lower energy consumption for sole encryption and authentication, respectively. Therefore, our proposed framework offers significant energy improvement over its standard counterpart for symmetric key primitives.

\textbf{Discussions: }We would like to note that our experiments were performed on the drone processor while flight control codes were inactive. When these codes are active, it would potentially increase the run-time of both frameworks. Since the impact of such a slow-down would be similar, we believe the improvements of our framework will remain. More importantly, the message throughput that both framework can support will decrease. Considering that the standard framework is already struggling meeting with the requirements of time-critical use-cases with current configurations, it will suffer more from this decrease in throughput. Therefore, we believe that, in a real-life deployment, the benefits of our improved framework would even increase and our framework can meet the needs of stringent requirements of real-time applications.

\section{Conclusion}\label{sec:Conclusion}
In this paper, to the best our knowledge, we proposed the first open-source low energy cryptographic framework tailored for small aerial drones with an in-depth energy consumption analysis. Our framework integrates algorithmic optimizations to improve the performance of standard PKC techniques, all supported with light-weight symmetric ciphers. We have implemented and deployed our optimized framework on Crazyflie 2.0, and compared its performance with the standard techniques. Our experiments confirmed that our improved framework offers up-to 35$\times$ higher speed and better energy efficiency compared with its standard counterpart. Therefore, our improved framework can meet some of the stringent energy and delay requirements of aerial drone networks with only a small impact on the battery life and end-to-end delay. We open-source our cryptographic framework for public testing, improvement and adoption purposes. \\
 
\textbf{Acknowledgment. }We would like to thank the anonymous
reviewers for their insightful comments and suggestions. This
work is supported by NSF CAREER Award CNS-1652389.

\bibliographystyle{IEEEtran}
\bibliography{../../../Cryptoetc/crypto-etc,tex_sig,../../../Cryptoetc/crypto-etc}

\end{document}